\documentclass[%
 reprint,
 showpacs,preprintnumbers,
 amsmath,amssymb,
 aps,
]{revtex4-1}

\usepackage{graphicx}
\usepackage{dcolumn}
\usepackage{bm}
\usepackage{CJK}

\begin{document}

\begin{CJK*}{GBK}{song}


\title{Spatiotemporal Intermittency on the Growing Surface of Coupled Sandpiles}
%
\author{Lei Liu}
\email{liulei@mail.iap.ac.cn}
%
\author{Fei Hu}

\affiliation{
 LAPC, Institute of Atmospheric Physics, Chinese Academy of
 Sciences, Beijing, China
}%

\date{\today}

\begin{abstract}
The surface of conservative coupled sandpiles in the self-organized cooperative critical state is found to exhibit intermittency in both time and space. The spatiotemporal intermittent structure is also found to be a multifractal. The probability density of sand grain heights on the surface is an asymptotic power law but with an exponential cut-off. The power spectra of the time series of sand grain heights show a diversity of low-frequency components over different sites on the surface and also over different ensemble samples. This means that the long-term prediction according to the nearby observations and the history experiences is very difficult in the world of coupled sandpiles. Unlike the low-frequency spectra, the high-frequency spectra seem to obey a universal $f^{-2}$ law.
\end{abstract}

\maketitle
\end{CJK*}

In 1980s, Bak, Tang and Wiesenfeld proposed the self-organized criticality (SOC) theory to explain the ubiquitous power laws in nature \cite{bak1,bak2,bak3}. This theory is simply illustrated by a {\it single} sandpile model (BTW model). Recently, we find a new kind of self-organized critical state by coupling two BTW sandpiles \cite{lh14}. We call it self-organized cooperative criticality (SOCC) because each sandpile in this model is not critical but they cooperate to be critical. In the SOCC state, we find that the distributions of the avalanche size and the lifetime exhibit the same finite-size scalings as those in the BTW sandpile. This conclusion has been strictly proved to be true \cite{dandekar}. However, there is a remarkable difference between the two kinds of criticality. In the BTW model, after each avalanche the negative value of local height gradient (for simplicity, this term is replaced by the ``local height'' in the following text) at any site will not exceed the critical threshold $z_c$ because once the local height exceeds $z_c$ an avalanche will be trigged again. However, in one of the coupled sandpiles the local height at each site may exceed $z_c$ and fluctuates irregularly in space and time because the trigger of avalanches is defined by the two sandpiles. The difference between the BTW model and our model is significance if we turn our eyes upon the natural world. For example, although the BTW model can explain the scaling behavior of earthquakes (avalanches of sandpile) successfully \cite{gr44,bak4}, it can not explain the very irregular configurations of the earth's surface (configuration of sandpile after avalanches). The irregular fluctuations in space and time is called the spatiotemporal intermittency here. In this paper, we will exhibit the quantitative features of spatiotemporal intermittency in the conservative coupled sandpiles. The results are expected to be used as fingerprints to search the SOCC state in real systems.

The conservative coupled sandpiles are composed of two BTW sandpiles which are coupled in such a way that, {\it if and only if} both local heights at the same coordinate of the coupled sandpiles (denoted by $z_1$ and $z_2$) are greater than the critical threshold $z_c$ (equaling 3 in the following simulations), the sand grains then tumble down as
\begin{equation}
\label{eq.1}
z_{i}(x,y)\rightarrow z_{i}(x,y)-4, \ \ {\rm for} \ \ i=1,2
\end{equation}
and the tumbling down grains are transported to the nearest-neighbor grids,
\begin{eqnarray}
\label{eq.2}
z_{i}(x\pm 1,y) & \rightarrow & z_{i}(x\pm 1,y)+1,\\
\label{eq.3}
z_{i}(x,y\pm 1) & \rightarrow & z_{i}(x,y\pm 1)+1.
\end{eqnarray}
Note that the coordinate of the upper left corner of the sandpile is set to $(1,1)$ and the coordinate of the lower right corner is set to $(N,N)$ in this paper. The coordinates range as $1\leq x\leq N$ and $1\leq y\leq N$. The coupled avalanche dynamics are inspired by the coupling phenomena in many real systems. For example, in our daily life you may have considered to migrate to a bigger city for a better job. Migration to a bigger city is a movement along the population gradient. Here, someone's final choice is determined by the equilibrium between the population gradient and other gradients, such as the job gradients, the education gradients and even shopping gradients. In other words, the dynamics of migration is coupled with other dynamics. In fact, similar coupling also exists in the natural systems. A well-known example is the cross effect in the transport arising in a mixture if both the concentrations and the temperature are non-uniform over the system \cite{gm84}.

\begin{figure}[hbtp]
\includegraphics[width=20pc]{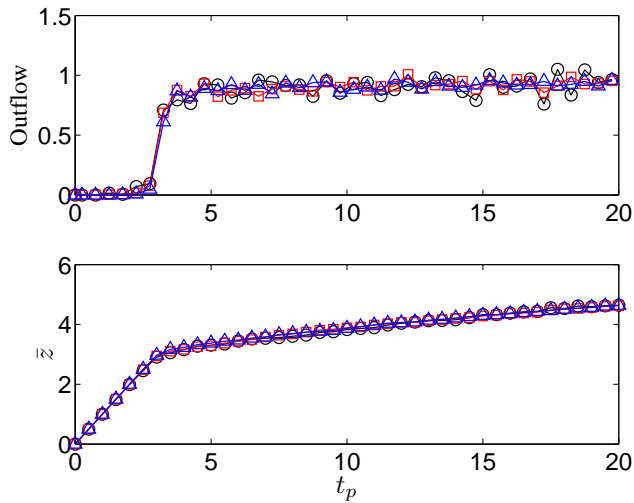}
\caption{Time evolutions of the coarse-grained outflow (top) and the local height averaged over the whole sandpile (bottom) for different sandpile sizes (circles for $N=20$, squares for $N=40$ and triangles for $N=60$).}
\label{fig.1}
\end{figure}

Starting from two sandpiles with the same size of $N$ and zero local height in the whole sandpiles, the critical states are built by adding one sand grain randomly to each sandpile simultaneously at every turn. After adding, the two coupled sandpiles are allowed to tumble down (if needed) according to Eqs.~(\ref{eq.1}), (\ref{eq.2}) and (\ref{eq.3}). This model is conservative except in the boundary where tumbled sand grains can flow out of the sandpile. Once the sandpiles become calm, another sand grain is added to each sandpile. Because of the symmetry of the two coupled sandpiles, we just show the statistics of one of them in the following analysis.

With the increase of number of adding grains $n$, a critical state will be built in the coupled sandpiles where the outflow will fluctuate around a constant value and the distributions of avalanche size and lifetime will follow the same power laws as those in the classic BTW sandpile. However, in the conservative coupled sandpile the local height averaged over the whole sandpile $\bar{z}$ will continue to increase with $n$ in the critical state while in the BTW sandpile it fluctuates around a constant \cite{lh14,jensen}. It means that the surface of coupled sandpiles shaped by the avalanches is growing with the elapse of time. One can image that in larger sandpiles more adding grains are needed than in smaller sandpiles to form a surface with a defined value of $\bar{z}$. Thus, the number of adding grains is not convenient to compare the degrees of growth with different sandpile sizes. We define the growth time $t_p$ to solve this problem
\begin{equation}
\label{eq.4}
t_{p}=\frac{n}{N^2},
\end{equation}
where $N$ is the sandpile size and $n$ is the number of adding grains. By using $t_p$, one can see that the curves of outflow measured with different sandpile sizes almost overlap each other (see the top plane of Fig.~\ref{fig.1}). The outflow begins to saturate at $t_p\approx 4$ where the sandpiles begin to be critical. By using $t_p$, the curves of the averaged local height $\bar{z}$ also overlap each other (see the bottom plane of fig.~\ref{fig.1}). It means that at the same growth time $t_p$ the degree of growth will be the same whatever the sandpile sizes are. In the following sections, we will show that $t_p$ can also be used to scale other variables into universal functions.

\begin{figure}[t]
\includegraphics[width=20pc]{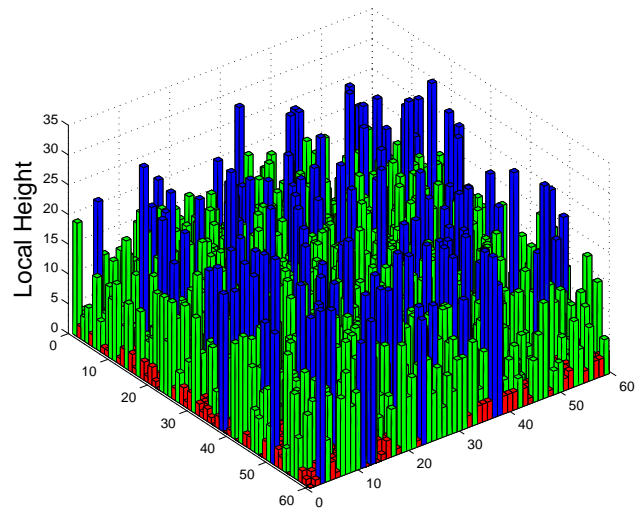}
\caption{A snapshot of the local heights $z$ on a SOCC sandpile with $N=60$ (blue bars for $z>20$, green bars for $3<z \leq 20$ and red bars for $z\leq 3$ ). The growth time $t_p=50$.}
\label{fig.2}
\end{figure}

{\bf Intermittency in space.} If the local height at a coordinate in one of coupled sandpiles is smaller than $z_c$, the local height at the same coordinate in another sandpile could be accumulated to be greater than $z_c$. This will lead to the spatial intermittency on the surface of coupled sandpiles (see Fig.~\ref{fig.2} for example). The quantitative features of spatial intermittency in the SOCC states, including the probability density functions of local heights and the sequence of mass exponent, are analyzed as follows.

\begin{figure}[t]
\includegraphics[width=20pc]{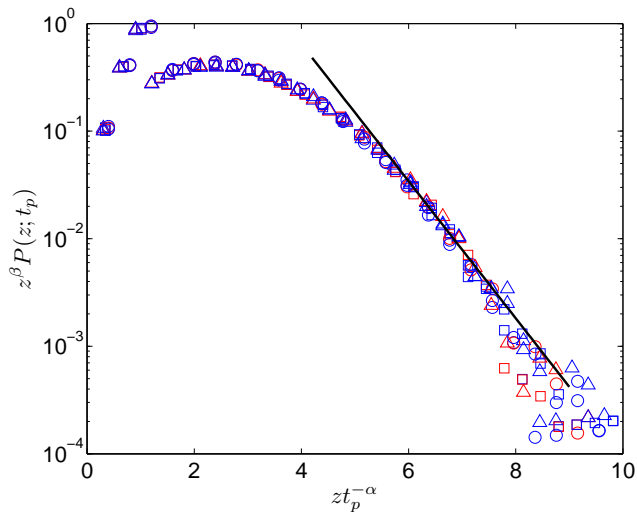}
\caption{Probability of density functions of local heights $P(z;t_p)$ for different sandpile sizes (black points for $N=20$, red points for $N=40$ and blue points for $N=60$) and growth times (circles for $t_p=10$, squares for $t_p=15$ and triangles for $t_p=20$) in the SOCC state. The slope of line is about $-1.5$.}
\label{fig.3}
\end{figure}

The probability density functions of local heights are calculated for different sandpile sizes and different growth times. In order to improve statistics, we generate many pairs of coupled sandpiles for each sandpile size ($500$ pairs for $N=20$, $126$ pairs for $N=40$ and $56$ pairs for $N=60$). We find that the probability density functions $P(z;t_p)$ behave as a universal function whatever the sandpile sizes are (see Fig.~\ref{fig.3}). The universal function is
\begin{equation}
\label{eq.5}
P(z;t_p)=z^{-\beta}f\left(\frac{z}{t^{\alpha}_p}\right)
\end{equation}
where $\alpha\approx 0.4$ and $\beta\approx 1.1$. When $zt^{-a}_p$ is very large, $P(z;t_p)$ will approach to a power law but with an exponential cut-off,
\begin{equation}
\label{eq.6}
P(z;t_p)\sim z^{-\beta}\exp\left(-\frac{c_1}{t^\alpha_p}z\right),
\end{equation}
where $c_1\approx 1.5$.

The success of the BTW sandpile model may give us a misleading impression that complex systems always follow asymptotic power-law distributions such as the distributions of earthquake sizes (the Gutenberg-Richter law \cite{gr44}), the incomes (the Pareto's law \cite{pareto}) and the word frequency (the Zipf's law \cite{zipf}). However, there are many other complex systems which do not follow the asymptotic power laws \cite{newman1}. In fact, the asymptotic power law with an exponential cut-off as one kind of the non-power-law distributions are observed in many systems. Examples include the turbulent wind fluctuations in the atmospheric boundary layer \cite{liu1,liu2}, the complex biological \cite{jmbo01} and social \cite{newman2} networks. Equation (\ref{eq.6}) shows that the SOCC could be an alternative mechanism of this kind of non-power-law distributions.

\begin{figure}[hbtp]
\centering
\includegraphics[width=20pc]{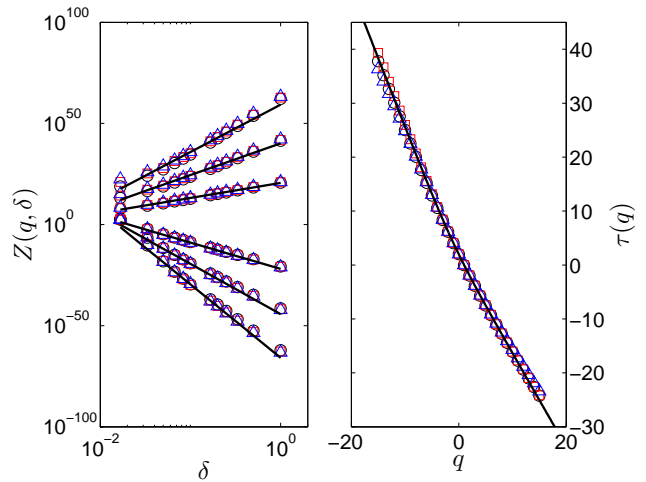}
\caption{(Left) The sum of $q$-th order moments of probability of local height $Z(q,\delta)$ as a function of box length $\delta$. Lines show that the data with different values of $q$ can be well fitted by the power law when $\delta$ is not very large. From the bottom line to the top line, the order $q=-15, -10, -5, 5, 10, 15$. In this plot, the side length of the square sandpile is set to $1$ and the sandpile size $N=60$. (Right) The sequence of mass exponents $\tau(q)$ as a function of moment order $q$. The line is calculated from Eq.~(\ref{eq.10}). In both panels, circles, squares and triangles denote the results for $t_p=10$, $t_p=15$ and $t_p=20$ respectively.}
\label{fig.4}
\end{figure}

The sequence of mass exponents is a useful tool to detect and describe the multifractal structure. Let us cover the whole sandpile with boxes of length $\delta$ and calculate the probability of local height in the $i$-th box
\begin{equation}
\label{eq.8}
\mu_i=\frac{z_i}{\sum^m_{i=1}z_i},
\end{equation}
where $m$ is the least number of boxes needed to cover the sandpile. If the sum of $q$-th moments of $\mu_i$ has the form
\begin{equation}
\label{eq.9}
Z(q,\delta)=\sum^m_{i=1}\mu^q_i \sim \delta^{-\tau(q)}
\end{equation}
when $\delta\rightarrow 0$ and if the power index $\tau(q)$ is a non-linear function of $q$, the surface of sandpile is then a multifractal \cite{feder}. Equation (\ref{eq.9}) is tested to be true for different growth times in the SOCC state (see the left panel of Fig.~\ref{fig.4}). The mass sequence $\tau(q)$ is also found to be a non-linear function of $q$ and can be fitted well with the mass exponents of 2D binomial multiplicative cascade process \cite{cheng05},
\begin{equation}
\label{eq.10}
\tau(q)=\frac{\ln\left(a^q+b^q+c^q+d^q\right)}{\ln2},
\end{equation}
where $a=0.3$, $b=0.3$, $c=0.23$ and $a+b+c+d\equiv 1$ (see the right panel of Fig.~\ref{fig.4}). The results show that the surface of coupled sandpiles in the SOCC state is a multifractal and its self-similarity feature will not change with $t_p$.

\begin{figure}[htbp]
\includegraphics[width=20pc]{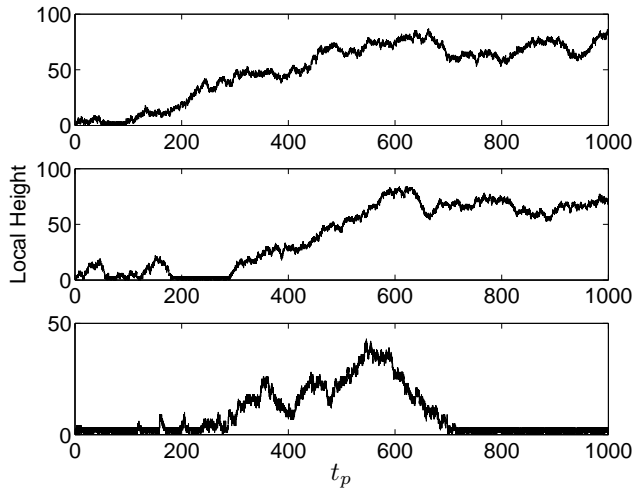}
\caption{Time series of local height at sites $(9,10)$ (top), $(6,5)$ (middle) and $(6,4)$ (bottom). The sandpile size $N=10$. Note that the time series at the same site will change with the ensemble sample of coupled sandpiles. The time series shown here are from a possible ensemble sample.}
\label{fig.5}
\end{figure}

{\bf Intermittency in time.} The time series of local height at different sites are shown in Fig.~\ref{fig.5}. Many interesting features of coupled sandpiles are shown in this figure. First, the coupled sandpiles have the complicate intermittent patterns in time even the sandpile size $N$ is not very large. The time evolutions of local height with a size of only $N=10$ show an variety of patterns (see Fig.~\ref{fig.5}). Second, adjacent sites could have dramatically different patterns of time evolution. For example, one can see that the dramatic difference between the time evolutions at adjacent sites $(6,5)$ and $(6,4)$ (see the middle and bottom planes of Fig.~\ref{fig.5}). This means that we can not make a good prediction of nearby sites according to the local experiences in the coupled sandpiles. Third, sites that are separated further from each other may have a similar pattern of time evolution. For example, in the top and middle planes of Fig.~\ref{fig.5}, one can see that both time series at the sites $(9,10)$ and $(6,5)$ have a long-term increasing trend first and then saturate at $t_p\approx 600$. We are not sure whether this similarity is related to some physical connection or just a coincidence.

\begin{figure}[htbp]
\centering
\includegraphics[width=20pc]{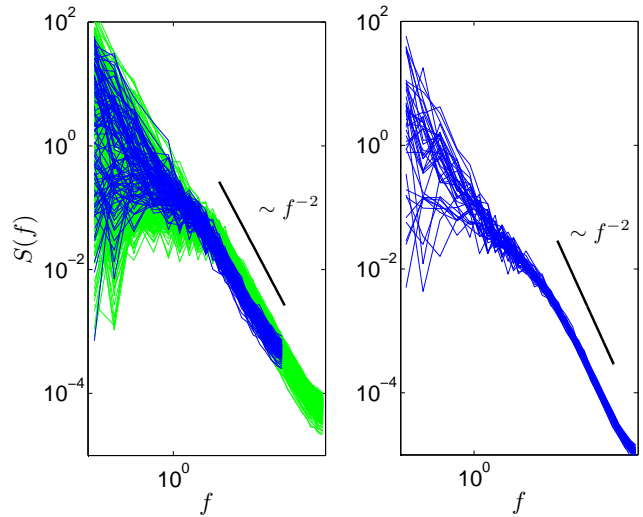}
\caption{(Left) Bin-averaged spectra at all sites on the SOCC sandpile. Green lines are for the sandpile size $N=20$ and blue lines are for $N=10$. (Right) Bin-averaged spectra at the site $(20,20)$ on the SOCC sandpile with a sandpile size of $N=40$. Spectra of $50$ ensemble samples are shown in this plot.}
\label{fig.6}
\end{figure}

The power spectra $S(f)$ of the time series of local height are analyzed at every site in the SOCC states. In the analysis, the sampling interval is set to $\triangle t_p=1/N^2$. Results show that the high-frequency spectra are similar for all sites and behave as a power law,
\begin{equation}
\label{eq.11}
S(f)\sim f^{-\xi},
\end{equation}
where $\xi\approx -2$ (see the left panel of Fig.~\ref{fig.6}). This seems to be true for different sandpile sizes. However, the low-frequency spectra are very different. This means that the long-term variations in the time series of height are diverse over the sandpile which is a reflection of the spatial intermittency. We generate many pairs of coupled sandpiles with a defined sandpile size and then analyze the spectra at a defined coordinate. It is found that the high-frequency spectra can also be described by a power law of Eq.~(\ref{eq.11}) for each ensemble sample. However, the low-frequency is also diverse over different ensemble samples (see the right panel of Fig.~\ref{fig.6}). This means that the randomness will severely affect the long-term variations in the coupled sandpiles. Thus, it will be very difficult to make a long-term prediction according to the history experience in the coupled sandpiles.

In conclusion, we have analyzed many quantitative features of spatiotemporal intermittency on the growing surface of coupled sandpiles in the SOCC state including the probability density functions of local height, the sequence of mass exponent and the spectra. It is found that these features exhibit some universal regularities, although the surface of coupled sandpiles seems to be very complicated in space and time. These findings may help us to diagnose SOCC states in real systems.

\begin{acknowledgments}
Part of this work is finished when one of the author Lei Liu visited the Earth System Physics section (ESP) of International Center of Theoretical Physics (ICTP). He is grateful for the hospitality of ESP, ICTP. The work is also supported by the National Nature Science Foundation of China (Grant No. 41105005).
\end{acknowledgments}

\bibliography{liuhu_manuscript}

\end{document}